\documentclass{ws-p9-75x6-50}
\begin{document}

   \title{On the self-consistence of electrodynamics in the early universe}


   \author{V. A. De Lorenci, R. Klippert, M. Novello and J. M. Salim}         
\address{CBPF, Brazilian Center for Research in Physics, 
Rua Dr.\ Xavier Sigaud 150 LAFEX, Urca - Rio de Janeiro, 22290-180 Brazil\\
e-mail: klippert@lafex.cbpf.br}

\maketitle

\abstracts{
The issue of a self-consistent solution of Maxwell-Einstein equations 
achieves a very simple form when all quantum effects are neglected 
but a weak vacuum polarization due to an external magnetic field 
is taken into account.  From a semi-classical point of view 
this means to deal with an appropriate limit of the one-loop 
effective Lagrangian for electrodynamics.  When the corresponding 
stress-energy tensor is considered as a source of the gravitational 
field a surprisingly bouncing behavior is obtained.  
The present toy model leads to important new features 
which should have taken place in the early universe.  
}

\section{Introduction}
\label{intro}

Classical electrodynamics provides results in very good agreement 
with cosmic observations from now up to the epoch near the end 
of the radiation era.  During the radiation dominated epoch, however, 
it is commonly adopted some inflationary mechanism, whose origin 
relies on quantum theory: inflatons, strings, membranes, etc.  
With this tool one describes the early universe matter content as an 
ionized plasma together with a large scale magnetic field,\cite{Adler} 
whose origin may rest on cosmic strings,\cite{Dimopoulos} 
pseudo Goldstone bosons\cite{Garretson} or other sources.  
At the GUT scale this magnetic field may reach very high intensities, 
$H\sim 10^{40}\,Gauss$ or even stronger.\cite{Heyl}  
Fields of this strength by far exceed the limit 
$H_c\approx 4.4\cdot 10^{13}\,Gauss$ 
beyond which QED vacuum polarization must be considered.  
One-loop effective potentials were thus calculated 
for a background magnetic field,\cite{Heisenberg,Schwinger,Reuter} 
the influence of them on the primordial magnetohydrodynamics 
being recently proposed.\cite{Berera}  

Nevertheless, no attempt to analyze the back-reaction 
of such corrections on the gravitational field itself was devised.  
In this vein, we will here limit our considerations to a toy model 
which describes the weak field limit of the 
one-loop zero temperature effective Lagrangian of QED 
driven by an external magnetic field.\cite{Reuter}  
The simplicity of the model allows its full solvability, 
yielding for a Friedmann-Robertson-Walker (FRW) spacetime 
a bouncing solution.  A natural upper bound for the magnetic field 
then arises, thus providing its mathematical consistence 
({\em i.e.}, the existence of a non-singular magnetic field 
defined throughout the whole history of the universe).  The rather 
distinct behavior of the above solution as compared with its classical 
counterpart (for which an initial singularity is unavoidable) 
suggests that the semi-classical treatment of the matter content 
of the actual universe could provide a very powerful tool 
for dealing with a cosmic singularity (and all the difficulties 
of standard cosmology from it derived, as the horizon problem).  
Other gauge interactions could similarly be considered, 
the cosmological relevance of them occurring as well at energy scales 
for which the Planck mass can still be neglected.  

In section \ref{Maxwell} we trace the origin of cosmic singularities 
in the most simple case of a FRW universe in the radiation era.  
It also presents the required spatial mean value algorithm 
to make such a model consistent with its own isotropy.  
Section \ref{Quantum} applies this same procedure to the weak field 
limit of QED effective Lagrangian, whose dynamics is obtained 
and solved in section \ref{Motion}.  Finally, section \ref{Matter} 
shows that classical ultra-relativistic matter fields 
(being the extreme limit of fast moving ions of the primordial plasma) 
cannot modify the regularity of the above referred solution.  

\section{Einstein-Maxwell singular universes}
\label{Maxwell}

Classical Maxwell electrodynamics leads to singular universes.
In a FRW framework this is a direct consequence 
of the singularity theorems,\cite{Penrose} which follows 
in this case from the exam of the energy conservation law and 
the Raychaudhuri equation for the Hubble expansion parameter $\theta$.
Let us set for the line element the form
\begin{equation}
\label{ds2}
ds^2 = c^2\,dt^2 - A^2(t)\,d\sigma^2.
\end{equation}
The 3-dimensional surface of homogeneity is orthogonal to a 
fundamental class of observers endowed with four-velocity 
vector field $v^{\mu} \doteq \delta^{\mu}_{t}$.  
In terms of the scale-factor $A(t)$, 
the expansion parameter is defined as 
\begin{equation}
\label{defTheta}
\theta \doteq 3\, \dot{A}/A,
\end{equation}
where a `dot' means partial derivative with respect to time 
({\em i.e.}, Lie derivative with respect to the velocity $v$).

For a perfect fluid with energy density $\rho$ and pressure $p$ 
the energy conservation law and the Raychaudhuri equation 
assume respectively the form\footnote{We will restrict ourselves 
throughout to the exam of the Euclidean section case.} 
\begin{equation}
\dot{\rho} + (\rho + p)\theta = 0,
\label{dotRho}
\end{equation}
\begin{equation}
\dot{\theta} + \frac{1}{3}\,\theta^2 = -\frac{\kappa}{2} (\rho+3p),
\label{dotTheta}
\end{equation}
in which $\kappa\doteq 8\pi{\rm G}/c^4$ 
is the Einstein gravitational constant.  
Equations (\ref{dotRho}) and (\ref{dotTheta}) do admit a first integral 
\begin{equation}
\label{constraint}
\theta^2 = 3\,\kappa\,\rho.
\end{equation}
Equations (\ref{constraint}) and (\ref{dotTheta}), 
which also imply (\ref{dotRho}) as well, can be identified 
with the timelike and the trace components of Einstein equations.

Since the natural spatial sections of FRW geometry are isotropic, 
electromagnetic fields can generate such universe 
only after a suitable spatial average be performed \cite{Hindmarth}.  
The standard procedure\cite{Tolman} is just to set%
\footnote{We make use of Gaussian Cartesian coordinates.  
Latin indices run in the spatial range $(x,\,y,\,z)$, 
while Greek indices run in the spacetime range $(t,\,x,\,y,\,z)$.} 
for the electric field $E_{i}$ and the magnetic field $H_{i}$ 
the following mean values: 
\begin{eqnarray}
<\,E_{i}\,> &=& 0,
\label{meanE}\\[1ex]
<\,H_{i}\,> &=& 0,
\label{meanH}\\[1ex]
<\,E_{i}\,E_{j}\,> &=& -\, \frac{1}{3} E^2 \,g_{ij},
\label{meanE2}\\[1ex]
<\,H_{i}\, H_{j}\,> &=&  -\, \frac{1}{3} H^2 \,g_{ij},
\label{meanH2}\\[1ex]
<\,E_{i}\, H_{j}\,> &=& 0.
\label{meanEH}
\end{eqnarray}
Here $E^2$ and $H^2$ are both nonnegative functions of time%
\footnote{They are not scalars, however, but depend on the 
set of coordinates, as far as expression (\ref{Ref-1}) 
is not a tensor definition but if $X$ is a scalar.}, 
and we denote by angular brackets the volume spatial average 
({\em e.g.}, $<X>$ represents the volume average 
of the arbitrary quantity $X$) for a given instant of time $t$, 
\begin{equation}
\label{Ref-1}
<\,X\,>\doteq\lim_{V\rightarrow V_o}\frac{1}{V}\int X\,d^3x^i,
\end{equation}
where $V=\int d^3x^i$ with $x^i\in\sigma$ being spatial coordinates, 
and $V_o$ stands for the time dependent volume of the whole space 
(which is finite for the closed section).  

The canonical stress-energy tensor 
associated with Maxwell Lagrangian is given by%
\footnote{We use Heaviside non-rationalized units throughout.}
\begin{equation}
T_{\mu\nu} = F_{\mu\,\alpha}\,F^\alpha{}_{\nu} 
+ \frac{1}{4} \,F \,g_{\mu\nu}, 
\label{TMaxwell}
\end{equation}
in which $F\doteq F_{\mu\nu}\,F^{\mu\nu}=2(H^2-E^2)$.  
Equations (\ref{meanE})--(\ref{meanEH}) imply
\begin{equation}
<\,F_{\mu\alpha}\,F^\alpha{}_\nu\,> = \frac{2}{3} (E^2 + H^2) 
v_{\mu}\,v_{\nu} + \frac{1}{3} (E^2 - 2\,H^2)\,g_{\mu\nu}.
\end{equation}
Using this result into the expression (\ref{TMaxwell}) of the 
stress-energy tensor, it follows that its average value 
$<T_{\mu\nu}>$ reduces to a perfect fluid configuration 
\begin{equation}
<T_{\mu\nu}>=(\rho_\gamma+p_\gamma)\,v_{\mu}\,v_{\nu}
-p_\gamma\,g_{\mu\nu},
\label{fluid}
\end{equation}
with energy density 
\begin{equation}
\label{RhoMaxwell}
\rho_\gamma = \frac{1}{2}\,(E^2 + H^2),
\end{equation}
and pressure 
\begin{equation}
\label{stateMaxwell}
p_\gamma = \frac{1}{3}\,\rho_\gamma.
\end{equation}
The fact that both the energy density (\ref{RhoMaxwell}) 
and the pressure (\ref{stateMaxwell}) are nonnegative for all time 
immediately yields the singular nature of classical FRW universes, 
as can be seen from Raychaudhuri equation (\ref{dotTheta}).  
In more precise words, Einstein field equations 
for the above energy-momentum configuration 
yield for the scale-factor the typical behavior 
\begin{equation}
\label{A(t)Maxwell}
A(t)\;\sim\;t^{1/2}.
\end{equation}

All this is standard and well-known.
However, near the maximum condensation era, classical Maxwell 
equations do not provide a correct description of electrodynamics. 
Instead, one should consider its quantum corrections. 

\section{Quantum corrections of QED at the radiation era}
\label{Quantum}

The effective action for electrodynamics 
due to one-loop quantum corrections 
was originally calculated by Heisenberg and Euler.\cite{Heisenberg}  
The development of this work to a gauge-invariant formulation 
is due to Schwinger.\cite{Schwinger}  
We present here only the first order calculation 
for the effective Lagrangian density
\begin{equation}
L=-\,\frac{1}{4}\,F+\frac{\mu}{4}\,F^2+\frac{7}{16}\,\mu\,G^2,
\label{Euler}
\end{equation}
in which $G\doteq F^{\star}_{\mu\nu}\,F^{\mu\nu}
=\frac{1}{2}\eta_{\alpha\beta\mu\nu}F^{\alpha\beta}F^{\mu\nu}
=-4\vec{E}\cdot\vec{H}$, with 
$F^\star_{\mu\nu}\doteq
\frac{1}{2}\eta_{\mu\nu\alpha\beta}F^{\alpha\beta}$.  
By $\eta_{\mu\nu\alpha\beta}$ we denote the Levi-Civita skew tensor, 
and 
\begin{equation}
\label{mu}
\mu\doteq\frac{2}{45}\,\alpha^2
\left(\frac{\hbar}{m_{e}\,c}\right)^3\frac{1}{m_{e}\,c^2}
\approx 1.67\cdot 10^{-31}\,cm^3/erg,
\end{equation}
where $\alpha\approx 1/137$ is the fine-structure constant.  

Note that the homogeneous Lagrangian (\ref{Euler}) 
requires some kind of spatial average over large scales, 
as given in (\ref{meanE})--(\ref{meanEH}).  
If one intends to make similar calculations on smaller scales, 
then either the more complex non homogeneous effective QED Lagrangian%
\cite{Dunne} should be used 
or else some additional magnetohydrodynamical effect%
\cite{Thompson,Subramanian} should be devised in order to achieve 
correlation\cite{Jedamzik} at the desired scale.  

Treating such quantum correction as a mere effective contribution 
to classical field theory, 
the corresponding modified stress-energy tensor reads\cite{Klippert}
\begin{equation}
\label{Tmunu}
T_{\mu\nu}=-4\,L_F\,F_\mu\mbox{}^\alpha F_{\alpha\nu}
+ (L_{G}G-L)\,g_{\mu\nu},
\end{equation}
in which $L_F$ represents the partial derivative of the Lagrangian $L$ 
with respect to the invariant $F$, and similarly for the invariant $G$.  

Since we are interested mainly in the analysis of the behavior 
of this system in the early universe, 
for which the actual ponderable matter should be identified 
with a primordial ionized plasma\cite{Tajima,Giovannini,Campos}, 
we are led to limit our considerations here to the case in which 
only the average of the squared magnetic field $H^2$ survives%
\footnote{This is strictly true for a viscosity free ionized plasma.  
When plasma viscosities are considered 
the resulting mean squared electric field may be non zero, 
but it would still be much smaller than its magnetic counterpart.}.%
\cite{Dunne,Tajima,Giovannini,Joyce}  
This is formally equivalent to put $E^2=0$ in (\ref{meanE2}).  
Let us see what the consequences of this result are.

\subsection{Equation of Motion and Energy Distribution}
\label{Motion}

Since the average procedure is independent from the equations 
of motion of the electromagnetic field we can use the above 
formulae (\ref{meanE})--(\ref{meanEH}) with $E^2=0$ to arrive at a similar 
expression as (\ref{fluid}) for the averaged stress-energy tensor, 
again identified with a perfect fluid with energy density 
$\rho_\gamma$ and pressure $p_\gamma$, which are given by 
\begin{eqnarray}
\rho_\gamma &=& \frac{1}{2} \, H^2 \,(1 - 2\,\mu\,H^2),
\label{rho}\\[1ex]
p_\gamma &=& \frac{1}{6} \,H^2 \,(1 - 10\,\mu\,H^2).
\label{P}
\end{eqnarray}
Since the averaged effective stress-energy tensor is not trace-free, 
the equation of state $p_\gamma = p_\gamma(\rho_\gamma)$
is no longer given by the Maxwell prescription (\ref{stateMaxwell}), 
but presents instead a new, quintessential,\cite{Caldwell} term 
which is proportional to the constant $\mu$ as 
\begin{equation}
\label{newstate}
p_\gamma = \frac{1}{3} \,\rho_\gamma - \frac{4}{3}\,\mu \,H^4.
\end{equation}
Equation (\ref{newstate}) can also be written in the form 
\begin{eqnarray}
p_\gamma &=& \frac{1}{3}\,\rho_\gamma-\frac{1}{6\mu}
\left[(1-2\mu\rho_\gamma)-\sqrt{1-4\mu\rho_\gamma}\right] \nonumber\\
&\approx& \frac{1}{3}(1-\mu\rho_\gamma)\rho_\gamma,
\label{Ref-2}
\end{eqnarray}
where the classical limit $\mu\rho_\gamma\ll 1$ was applied 
to obtain the last equality in (\ref{Ref-2}).

We note that, as $\mu$ is a positive constant, 
one could envisage the possibility 
that both the energy density and the pressure could become negative.  
We shall see below that this does not occur for the energy density, 
but it is precisely the case for the pressure.  
Specifically, there exists an inflationary epoch 
in the model of the universe presented here, 
for which the inertial energy condition ($\rho+p\ge 0$) is violated; 
the gravitational energy condition ($\rho+3p\ge 0$) 
is violated as well.  

Equations (\ref{dotRho}) and (\ref{constraint}) do encompass 
the whole dynamics.  Indeed, from (\ref{rho})--(\ref{P}) we have: 
\begin{equation}
\label{dotA}
H^2 \,\left( 1 - 4\,\mu\,H^2 \right) \,\left(\frac{\dot{H}}{H} + 
2\, \frac{\dot{A}}{A} \right)  = 0,
\end{equation}
and
\begin{equation}
\label{ddotA}
\frac{\dot{A}^2}{A^2}=\frac{\kappa}{6}\,H^2\,(1-2\,\mu\,H^2).
\end{equation}
Equation (\ref{dotA}) furnishes%
\footnote{We shall not consider here the $H^2=constant$ case, 
as it does not fit the present behavior of the actual universe.} 
\begin{equation}
H=\frac{H_o}{A^2},
\label{H->A}
\end{equation}
where $H_o$ is a nonnegative constant.  
With this result, (\ref{ddotA}) turns out to be an ordinary 
first order nonlinear differential equation for the scale-factor
\begin{equation}
\protect\label{eqA}
\dot{A}^2 = 
\frac{\kappa\,H_o^2}{6\,A^2} \left(1-\frac{2\,\mu\,H_o}{A^4}\right),
\end{equation}
whose solution is\footnote{We omit here an integration constant, 
which represents the origin of the time marks, 
by arbitrarily setting $\left.A\right|_{t=0}\equiv A_{\min}$.} 
\begin{equation}
\label{A(t)}
A^2 = H_{o} \,\sqrt{\frac{2}{3} \,(\kappa\,c^2\,t^2 +3\,\mu)}.
\end{equation}
Equation (\ref{H->A}) then yields the magnetic field as 
\begin{equation}
\protect\label{H(t)}
H^2 = \frac{3}{2}\,\frac{1}{\kappa\,c^2\,t^2 + 3\,\mu},
\end{equation}
whose maximum value 
\begin{equation}
\label{Hmax}
H_{max}=\frac{1}{\sqrt{2\,\mu}}\approx 6.14\cdot 10^{15}\,Gauss
\end{equation}
is much smaller than the typical values 
which occurr at the GUT scale.\cite{Heyl}  

\begin{figure}[htb]
\leavevmode
\centering
\epsfxsize=40ex
\epsfysize=40ex
\epsfbox{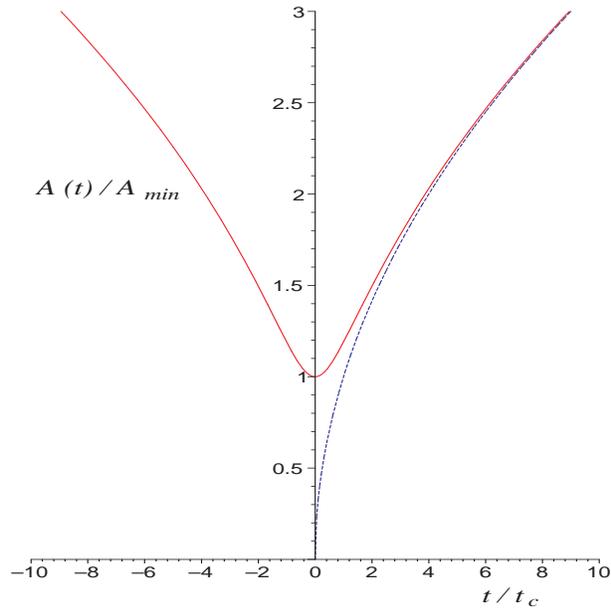}
\caption{Non-singular behavior of the scale-factor $A(t)$.  
$A_{min}$ is given from (\protect\ref{Amin}) 
and $t_c$ from (\protect\ref{tc}).}
\label{FigA(t)}
\end{figure}

Let us make some comments on this solution.  
First of all we realize it is a bouncing solution\footnote
{For alternative models with bouncing FRW solutions see, {\em e.g.},%
\cite{Ruzmaikina,Gurovich,Murphy,Bekenstein,Melnikov,Salim,Randibar,%
Novello,Balbinot,Elbaz,Alvarenga,Sautu,Dirar,Mannheim,Gasperini}.}, 
as displayed in Figure \ref{FigA(t)}, 
whose minimum ``radius'' $A_{min}$ is given by 
\begin{equation}
\label{Amin}
A^2_{min} = H_{o} \, \sqrt{2\,\mu}.
\end{equation}
The actual value of $A_{min}$ then depends on the constant $H_o$, 
which turns out to be the unique free parameter of the present model.  

\vspace{1.8cm}

\begin{figure}[htb]
\leavevmode
\centering
\epsfxsize=40ex
\epsfysize=40ex
\epsfbox{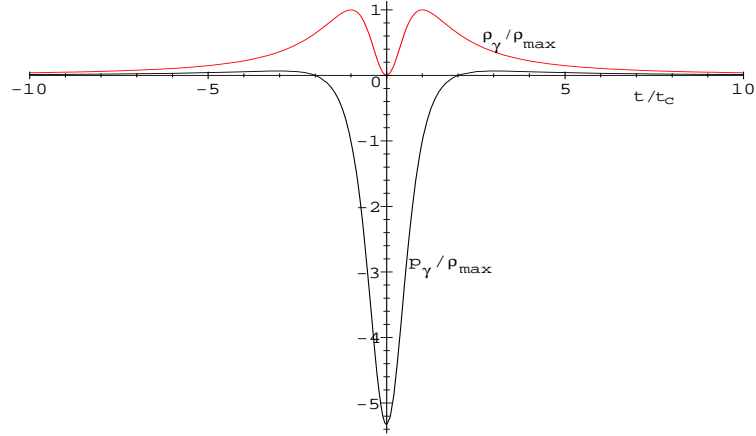}
\caption{Time dependence of the electromagnetic 
energy density $\rho_\gamma$ and pressure $p_\gamma$.  
$\rho_{max}$ is given from (\protect\ref{maxRho}) 
and $t_c$ from (\protect\ref{tc}).}
\label{FigP(t)}
\end{figure}

The energy density $\rho_\gamma$ attains its maximum value 
\begin{equation}
\label{maxRho}
\rho_{max} = \frac{1}{16\,\mu} \approx 3.75\cdot 10^{27}\,erg/cm^3
\end{equation}
at the instant $t=t_c$, where
\begin{equation}
\label{tc}
t_c = \frac{1}{c} \,\sqrt{\frac{3\,\mu}{\kappa}}
\approx 1.64\cdot 10^{-2}\,sec.
\end{equation}
For lower values of $t$ the energy density decreases, 
vanishing at $t=0$.  At the same time the pressure becomes 
highly negative, as displayed in Figure \ref{FigP(t)}.  
Only for times comparable to (or smaller than) $t_c$, which lies 
beyond the observational lower limit $t\ge t_n\sim 10^0\,sec$ 
provided by the nucleosynthesis \cite{Alvarenga}, 
the quantum effects are important.  
Indeed, solution (\ref{A(t)}) fits the standard expression 
(\ref{A(t)Maxwell}) of the Maxwell case at the limit of large times.  

The corresponding maximum of the temperature is given by 
\begin{equation}
\label{Tmax}
T_{max}=\left(\frac{c}{8\mu\sigma}\right)^{1/4},
\end{equation}
where $\sigma$ is the Steffan-Boltzmann constant.  
In energy units the chosen value of $\mu$ 
(which describes virtual pairs of {\bf electrons}) yields 
\begin{equation}
\label{Emax}
k_B T_{max} \approx 12.2 \, MeV.
\end{equation}
This result is just at the upper limit for which QED 
admits a perturbative expansion on the temperature-dependent 
coupling constant\cite{Ahmed} $e$, $T^{QED}_{max}\sim 10\,MeV$.  
We do not care of this point, since the vacuum polarization process 
we are interested in does not require that the virtual pairs $e^+e^-$ 
be produced in thermal equilibrium with the electromagnetic field 
which generates them.  
Indeed, standard calculations\cite{Mostepanenko} often suppose, as we did, 
that these virtual pairs are created mostly at zero temperature.

Temperatures of the order of (\ref{Emax}) can also induce 
virtual pairs of heavier particles, as the up quark.  
Since the right-hand side of (\ref{Tmax}) increases linearly 
with the virtual mass/charge ratio this would in fact lead to a cascade 
effect which would presumably end at the heaviest relevant particle: 
the top quark, whose mass is\cite{Caso} $m_t/m_e\approx 3.33\cdot 10^5$.  
If one considers this additional contribution 
the maximum thermal energy (\ref{Emax}) would then be shifted 
to about $k_B T_{max} \approx 2.70 \, TeV$, 
occurring near $t_c\approx 3.33\cdot 10^{-13}\,sec$, 
while $H_{max}\approx 3.02\cdot 10^{22}\,Gauss$.  

\section{Matter fields at the radiation era}
\label{Matter}

Let us make another comment concerning the influence 
of the presence of other matter fields in the universe. 
Beside photons there are plenty of other particles, 
and physics of the early universe deals with various sort of fields. 
In the standard framework they are treated in terms of a fluid 
with energy density $\rho_\nu$, which satisfies 
an ultrarelativistic equation of state $p_\nu=\rho_\nu/3$.  
Adding the contribution of this kind of matter to the averaged 
stress-energy tensor $<T_{\mu\nu}>$ of the electromagnetic field 
given by (\ref{fluid}) and (\ref{rho}) and (\ref{P})
it follows, as usual, that $\rho_\nu$ is proportional to 
the inverse of the fourth power of the scale-factor
\begin{equation}
\label{otherRho}
\rho_\nu = \rho_\nu^{(o)} A^{-4}.
\end{equation} 
This result allows us to treat such extra matter fields 
as a mere reparametrization of the constants $H_o$ and $\mu$ 
into $\hat{H}{}_o$ and $\hat{\mu}$, given by
\begin{eqnarray}
\label{newH}
\hat{H}{}_o^2 &=& H_o^2 + 2\, \rho_{\nu}^{(o)},\\[2ex]
\label{newMu}
\hat{\mu} &=& 
\left(\frac{H_o^2}{H_o^2+2\,\rho_\nu^{(o)}}\right)^2\,\mu.
\end{eqnarray}
The net effect of this is just to re-scale the value of $A_{min}$ as 
\begin{equation}
\label{newA}
\hat{A}{}_{min}=
\left(\frac{H_o^2}{H_o^2+2\rho_\nu^{(o)}}\right)^{1/4}A_{min}.
\end{equation}
Therefore, it turns out that the phenomenon of reversing the sign 
of the acceleration parameter $\dot\theta$ due to the high negative 
pressure of the magnetic field is not essentially modified 
by the presence of the ultrarelativistic gas.  
Only a fluid possessing energy density $\rho_{exotic}$ 
which scales as $\rho_{exotic} = \rho_{exotic}^{(o)} \, A^{-n}$ 
with $n\ge 8$ could be able to modify the above result.  
However, this seems to be a very unrealistic case \cite{Caldwell}.  

\section{Conclusions}
\label{Conclusions}

Heisenberg and Euler\cite{Heisenberg} have calculated the 
effective Lagrangian density to deal with the nonlinear 
electrodynamic effects induced by virtual electron-positron pairs.  
This is valid for frequencies $\nu\ll m_e\,c^2/h$.  
One should wonder if the use of such a correction 
in the framework of Einstein general relativity, as we did before, 
belongs to the above range of applicability.  
Let us write the on-shell Heisenberg-Euler effective Lagrangian 
(\ref{Euler}) in the form
\begin{equation}
\label{onshell}
L = -\, \frac{1}{2} H^2 \left(1 - 2\,\mu\,H^2 \right).
\end{equation}
The limit of validity of this expression consists in the range 
\begin{equation}
\label{rangeH}
1 - 2\,\mu\,H^2 \ge 0.
\end{equation}
In the history of the universe described above 
the spatially averaged magnetic field strength $H^2$ 
is globally regular, and is bounded from above at the precise value 
for which the equality in (\ref{rangeH}) holds, 
as seen from equation (\ref{rho}) and Figure \ref{FigP(t)}.  
Such equality is, however, an extreme limit for the weak field 
expansion which is assumed to hold from the very beginning.  
The involved series is convergent for all times 
but for the instant of maximum condensation, $t=0$.  
Even worser, the dropped terms in the effective Lagrangian 
(\ref{Euler}) are not negligible at $t=t_c$ also.  
Therefore, the present theory must be regarded as a toy model 
on cosmology, although being a regular and consistent solution 
of Maxwell-Einstein equations.  
Moreover, this model explicitly provides an example 
of the cosmological relevance of a semi-classical description 
of matter at the early universe.  

From what we have seen, the cosmic singularity of standard cosmology 
is linked with the strictly classical framework in which 
both the gravitational field and the matter content of the universe 
are treated.\cite{Kolb}  
While the temperature increases the quantum effects become important.  
The present paper thus analyzes the weak field limit of the one-loop 
zero temperature electrodynamics from a semi-classical approach%
\cite{Reuter} in the realm of a spatially homogeneous and isotropic 
cosmology driven by an external magnetic field.\cite{Berera}  
Instead of being conclusive, however, the regular solution (\ref{H(t)}) 
suggests that more accurate descriptions of matter fields at the 
early universe may provide as well a globally consistent solution.  
The scenario devised here deserves therefore further investigation.

\section*{Acknowledgements}
This work was partially supported by the Brazilian Agencies {\em
Conselho Nacional de Desenvolvimento Cient\'{\i}fico e Tecnol\'ogico} (CNPq), 
{\em Funda\c{c}\~ao de Amparo \`a Pesquisa do Estado do Rio de Janeiro} 
(FAPERJ) and {\em Funda\c{c}\~ao Coordena\c{c}\~ao de Aperfei\c{c}oamento 
de Pessoal de N\'\i vel Superior} (CAPES).



\begin{thebibliography}{88}
%
\bibitem{Adler}
S. L. Adler, {\em Ann.\ Phys}.\ {\bf 67}, 599 (1971).
%
\bibitem{Dimopoulos}
K. Dimopoulos, {\em Phys.\ Rev.\ D} {\bf 57} (8), 4629 (1998).
%
\bibitem{Garretson}
W. D. Garretson, G. B. Field and S. M. Carroll, 
{\em Phys.\ Rev.\ D} {\bf 46} (12), 5346 (1992).
%
\bibitem{Heyl}
J. S. Heyl and L. Hernquist, 
{\em Phys.\ Rev.\ D} {\bf 59} (4), 045005 (1999).
%
\bibitem{Heisenberg}
W. Heisenberg and H. Euler, {\em Z. Phys}.\ {\bf 98}, 714 (1936).
%
\bibitem{Schwinger}
J. Schwinger, {\em Phys.\ Rev}.\ {\bf 82} (5), 664 (1951).
%
\bibitem{Reuter}
W. Dittrich and M. Reuter, in {\em Effective Lagrangians 
in Quantum Electrodynamics} (Springer-Verlag, Berlin-Heidelberg, 1985).
%
\bibitem{Berera}
A. Berera, T. W. Kephart and S. D. Wick, 
{\em Phys.\ Rev.\ D} {\bf 59} (4), 043510 (1999).
%
\bibitem{Penrose}
S. W. Hawking and G. F. R. Ellis, 
in {\em The Large Scale Structure of Spacetime} 
(Cambridge University Press, Cambridge, 1973)
and references therein concerning the singularity theorems.
%
\bibitem{Hindmarth}
M. Hindmarth and A. Everett, {\em Phys.\ Rev.\ D} {\bf 58} 103505 (1998).
%
\bibitem{Tolman} 
R. C. Tolman and P. Ehrenfest, {\em Phys.\ Rev}.\ {\bf 36}, 1791 (1930). 
%
\bibitem{Dunne}
G. Dunne and T. Hall, {\em Phys.\ Rev.\ D} {\bf 58} 105022 (1998); 
G. Dunne, {\em Int.\ J. Mod.\ Phys.\ A} {\bf 12} (6), 1143 (1997).
%
\bibitem{Thompson}
C. Thompson and O. Blaes, {\em Phys.\ Rev.\ D} {\bf 57} (6), 3219 (1998).
%
\bibitem{Subramanian}
K. Subramanian and J. D. Barrow, {\em Phys.\ Rev.\ D} {\bf 58} 883502 (1998). 
%
\bibitem{Jedamzik}
K. Jedamzik, V. Jatalini\'c and A. V. Olinto, 
{\em Phys.\ Rev.\ D} {\bf 57} (6), 3264 (1998).
%
\bibitem{Klippert}
M. Novello, V. A. De Lorenci, J. M. Salim and R. KLippert, 
{\em Phys.\ Rev.\ D} {\bf 61} (4), 045001 (2000). 
%
\bibitem{Tajima}
T. Tajima, S. Cable, K. Shibata and R. M. Kulsrud, 
{\em Astrophys.\ J.} {\bf 390}, 309 (1992).
%
\bibitem{Giovannini}
M. Giovannini and M. Shaposhnikov, 
{\em Phys.\ Rev.\ D} {\bf 57} (4), 2186 (1998).
%
\bibitem{Campos}
A. Campos and B. L. Hu, {\em Phys.\ Rev.\ D} {\bf 58} 125021 (1998).
%
\bibitem{Joyce}
M. Joyce and M. Shaposhnikov, 
{\em Phys. Rev. Lett}.\ {\bf 79} (7), 1193 (1997).
%
\bibitem{Caldwell}
R. R. Caldwell, R. Dare and P. J. Steinhardt, 
{\em Phys.\ Rev.\ Lett}.\ {\bf 80} (8), 1582 (1998).
%
\bibitem{Ruzmaikina}
T. V. Ruzmaikina and A. A. Ruzmaikin, 
{\em Sov.\ Phys.\ JETP} {\bf 30}, 372 (1970).
%
\bibitem{Gurovich}
V. Ts.\ Gurovich, {\em Sov.\ Phys.\ Dokl}.\ {\bf 15}, 1105 (1971).
%
\bibitem{Murphy}
G. L. Murphy, {\em Phys.\ Rev.\ D} {\bf 8} (12), 4231 (1973).
%
\bibitem{Bekenstein}
J. D. Bekenstein, {\em Phys.\ Rev.\ D} {\bf 11} (8), 2072 (1975).
%
\bibitem{Melnikov}
V. N. Melnikov and S. V. Orlov, 
{\em Phys.\ Lett.\ A} {\bf 70}, 263 (1979).
%
\bibitem{Salim}
M. Novello and J. M. Salim, 
{\em Phys.\ Rev.\ D} {\bf 20}, 377 (1979).
%
\bibitem{Randibar}
S. Randibar-Daemi, A. Salam and J. Strathdee, 
{\em Phys.\ Lett.\ B} {\bf 135} (5,6), 388 (1984).
%
\bibitem{Novello}
M. Novello and H. Heitzmann, 
{\em Gen.\ Relat.\ Grav}.\ {\bf 16}, 535 (1984).
%
\bibitem{Balbinot}
R. Balbinot and J. C. Fabris, 
{\em Gen.\ Relat.\ Grav}.\ {\bf 23} (12), 1307 (1991).
%
\bibitem{Elbaz}
M. Novello, L. A. R. Oliveira, J. M. Salim and E. Elbaz, 
{\em Int.\ J. Mod.\ Phys.\ D} {\bf 1} (3 \& 4), 641 (1993).
%
\bibitem{Alvarenga}
F. G. Alvarenga and J.C. Fabris, 
{\em Gen.\ Relat.\ Grav}.\ {\bf 28} (6), 645 (1996).
%
\bibitem{Sautu}
J. C. Fabris, J. M. Salim and S. L. Sautu, 
{\em Mod.\ Phys.\ Lett.\ A} {\bf 13} (12), 953 (1998).
%
\bibitem{Dirar}
M. Dirar, A. El-tahir and M. H. Shaddad, 
{\em Mod.\ Phys.\ Lett.\ A} {\bf 13} (37), 3025 (1998).
%
\bibitem{Mannheim}
P. D. Mannheim, {\em Phys.\ Rev.\ D} {\bf 58} 103511 (1998).
%
\bibitem{Gasperini}
M. Gasperini, {\em Gen.\ Relat.\ Grav}.\ {\bf 30} (12), 1703 (1998).
%
\bibitem{Ahmed}
E. Ahmed and J. G. Taylor, 
{\em Gen.\ Relat.\ Grav}.\ {\bf 20} (4), 395 (1998).
%
\bibitem{Mostepanenko}
A. A. Grib, S. G. Mamayev and V. M. Mostepanenko, 
in {\em Vacuum Quantum Effects in Strong Fields} 
(Friedmann Laboratory Publishing, St.\ Petersburg, 1994).
%
\bibitem{Caso}
C. Caso {\em et al}, {\em The European Phys.\ J. C} {\bf 3}, 1 (1998).
%
\bibitem{Kolb}
E. W. Kolb and M. S. Turner, in {\em The Early Universe} 
(Addison Wesley, California, 1990).
%
\end{thebibliography}
\end{document}